\documentclass{article}
\usepackage{ijcai21}
\usepackage{times}
\usepackage{soul}
\usepackage{url}
\usepackage[hidelinks]{hyperref}
\usepackage[utf8]{inputenc}
\usepackage[small]{caption}
\usepackage{graphicx}
\usepackage{amsmath}
\usepackage{booktabs}
 \usepackage{algpseudocode}
\usepackage{multirow}
\usepackage{paralist}
\usepackage{enumitem}
\usepackage{subfig}
\usepackage{url}
\usepackage{float}
\usepackage[table]{xcolor}
\usepackage{balance}
\usepackage{adjustbox}
 \usepackage[ruled,lined,linesnumbered]{algorithm2e}
\usepackage{xcolor}

\usepackage{soul}
\usepackage{url}
\usepackage[hidelinks]{hyperref}
\usepackage[utf8]{inputenc}
\usepackage{graphicx}

\usepackage{amsmath}
\usepackage{booktabs}
\usepackage[compact]{titlesec}

\usepackage{capt-of}

\usepackage{capt-of}

\newcounter{subfloat}

\usepackage[ruled,lined,linesnumbered]{algorithm2e}
\def\BibTeX{{\rm B\kern-.05em{\sc i\kern-.025em b}\kern-.08emT\kern-.1667em\lower.7ex\hbox{E}\kern-.125emX}}

\title{A Rule Mining-Based Advanced Persistent Threats Detection System}
\author{
Sidahmed Benabderrahmane$^{1,2}$\footnote{Contact Author}\and
Ghita Berrada$^{3,4}$\and
James Cheney$^{1,5}$\And
Petko Valtchev$^6$\\
\affiliations
$^1$The University of Edinburgh, School of Informatics, Edinburgh, UK.\\
$^2$New York University, Computer Science Department.\\
$^3$King's College London, School of Population Health and Environmental Sciences, UK. \\
$^4$University of Manchester, School of Health Sciences, UK.\\
$^5$The Alan Turing Institute, UK. \\
$^6$Université du Québec à Montréal, CRIA, Montréal (QC), Canada.\\
\emails
sidahmed.benabderrahmane@gmail.com, ghita.berrada@kcl.ac.uk, jcheney@inf.ed.ac.uk, valtchev.petko@uqam.ca
}
\begin{document}
\maketitle
\begin{abstract}
Advanced persistent threats (APT) are stealthy cyber-attacks that are aimed at stealing valuable information from target organizations and tend to extend in time. 
Blocking all APTs is impossible, security experts caution, hence the importance of research on early detection and damage limitation. 
Whole-system provenance-tracking and provenance trace mining are considered promising as they can help find causal relationships between activities and flag suspicious event sequences as they occur.
We introduce an unsupervised method that exploits OS-independent features reflecting process activity to detect 
realistic APT-like attacks from provenance traces. 
Anomalous processes are ranked using both frequent and rare event associations learned from traces. Results are then presented as implications which, since interpretable, help leverage causality in explaining the detected anomalies. When evaluated on Transparent Computing program datasets (DARPA), our method outperformed competing approaches.
\end{abstract}
\section{Introduction}
``[I]n this world nothing can be said to be certain, except death and taxes.'', wrote Benjamin Franklin in 1789. To this list, one could add the increasingly common and headline-grabbing cyberattacks/security breaches~\cite{sebenius2021solarwinds,wang2018cybersecurity,huang2018powergrid},
most of which are so-called advanced persistent threats (APT).

APTs are long-running and stealthy cyberattacks where adversaries gain access to specific targets' systems, staying undetected in the system for as long as necessary to reach their goal (mostly stealing or corrupting sensitive data or damaging the target's critical systems). 
Such attacks are now ``part and parcel of doing business''~\cite{auty2015anatomy}, say experts and when not stopped early enough, inflict significant damage, particularly financial or reputational, to the victim. Preventing all such attacks is impossible~\cite{auty2015anatomy}, experts warn, so systems should be monitored continuously so as to detect APTs early and keep their damage to a minimum. 

With APTs mimicking normal user activity, such attacks cannot be detected with traditional means (e.g antivirus software, signature or system-policy-based techniques). Methods relying on system/event logs or audit trails typically fail as they generally only analyze short event/system call sequences, which not only makes them unable to properly model and capture long-term behavior patterns but also susceptible to evasion techniques. Recent work~\cite{park2012provenance,streamspot,han2018tapp,DBLP:journals/fgcs/BerradaCBMMTW20,unicorn} has suggested whole-system provenance-tracking and provenance trace mining as solutions better suited for APT detection: the richer contextual information of provenance would help identify causal relationships between system activities, allowing the detection of attack patterns (e.g data exfiltration) that usually go unnoticed with the usual perimeter defence-based or policy-driven tools~\cite{jewell2011host,zhang2012track,awad2016data,jenkinson2017applying}. 

There are, however, challenges to be overcome before mining provenance data that can fulfill its system security strengthening promises. There are also issues directly linked to the recording of provenance itself (e.g. level of provenance granularity, fault tolerance, trustworthiness and consistency of the recorded trace~\cite{jenkinson2017applying}). 
The more worrisome issue though is the ``needle in a haystack'' problem: the volume of recorded provenance traces is massive (each day of system activity results in one or more gigabytes of provenance traces, containing hundreds or thousands of processes) and anomalous system activity (if at all present) only constitutes a tiny fraction of the recorded traces. Added to this, the diversity of possible APT patterns and the unavailability of fully annotated data make it even more complex to uncover anomalous activity indicative of an ongoing APT.

In such a context, typical supervised learning techniques would be of limited use to detect APT patterns and only unsupervised learning techniques would pass muster (see section~\ref{subsec:datasets} on data imbalance). In operational security settings, the ready availability of actionable information is critical. Security analysts can easily recognize and forensically investigate suspicious system behavior (e.g. processes created or subverted by an attacker) when brought to their attention. However, having the analysts sift through the traces in their entirety, when as little as 0.004\% of the activity, if any at all, is suspicious, is hardly an efficient use of their time. 
In this paper, we investigate the key subproblem of quickly flagging unusual process activity that calls for further manual inspection. To tackle this problem, we summarize process activity through binary or categorical features (e.g kinds of events performed by a process, process executable name, IP addresses and ports accessed) and propose an anomaly detection method that scores individual processes based on whether they satisfy or violate association rules (both frequent and rare) extracted from process activity summaries. 

Central to our method and a key advantage of it is the use of association rules. Not only do the rules let us score processes and flag anomalies in a principled manner, but they also allow us to present results in a way that could be more easily understood and interpreted by experts/security analysts seeking to thoroughly investigate and gain a deeper understanding of attack patterns.
We evaluated our approach using provenance traces produced by the DARPA Transparent Computing (TC) program\footnote{https://www.darpa.mil/program/transparent-computing }. We set our association-based anomaly detector against an array of existing batch anomaly detectors and show it outperformed all of them. 

The remainder of the paper is as follows: Section~\ref{sec:rel} summarizes related work, while section~\ref{sec:arm} provides background on association rule mining. Next, section~\ref{sec:approach} introduces our approach and section~\ref{sec:results} discusses the experimental study and its outcome. We conclude with section~\ref{sec:concl}.
\section{Related Work}
\label{sec:rel}
Intrusion and malware detection methods follow two main approaches: misuse detection
(e.g.~\cite{kumar1994pattern}) and anomaly detection
(e.g. \cite{ji2016multi}).  Misuse detection
searches for events matching predefined signatures
and patterns. Related methods can only detect
attacks with known signature/patterns, hence are 
unsuitable for APT detection. By contrast, anomaly detection makes no assumption about the attack nature and just looks for activity that deviates from normal behavior
i.e. usually recorded on a specific host or network.  
Ahmed et
al.~\cite{ahmed2016survey} is a survey of the main network
 anomaly detection techniques, which it divides into four categories:
 classification-based, clustering-based, information theory-based and statistical.
Anomaly detection surveys~\cite{anomaly,graph-anomaly} typically distinguish approaches w.r.t. the type of data (categorical, continuous, semi-structured, etc.) they admit.
Among those, graph methods are the most relevant for our study, yet they typically work on graph formats of reduced expressiveness (e.g. undirected or unlabeled), whereas provenance graphs have rich structure (labels and properties on both nodes and edges). Existing anomaly detection approaches for provenance graphs rely on training with benign traces~\cite{streamspot}, require user-provided annotations~\cite{sleuth}, or assume highly regular background activity~\cite{winnower}. 

In parallel, a number of anomaly detection approaches have been designed for categorical
data~\cite{fpoutlier,outlierdegree,avf,krimp-ad,comprexpaper,upc}.
Some of them, e.g. OC$3$~\cite{krimp-ad,krimp} and CompreX~\cite{comprexpaper}, are based on the Minimum Description Length (MDL) principle~\cite{mdl}: the idea here is to preprocess the dataset by compressing it and then take the compressed
record size as its anomaly score, the underlying assumption being that infrequent/anomalous patterns are less efficiently compressed and result in higher sizes. UPC~\cite{upc} also relies on MDL (in combination with pattern mining) and is a two-pass approach that looks for a different kind of anomalies (\emph{class-2}) than CompreX. FPOF (Frequent Pattern Outlier Factor)~\cite{fpoutlier} is an itemset-mining method exploiting transaction outlierness: outliers (lower FPOF values) are transactions with fewer frequent patterns. 
Outlier-degree (OD)~\cite{outlierdegree}, uses both categorical and continuous variables whereby infrequent values would indicate outlier status. %
AVF (Attribute Value Frequency)~\cite{avf} 
computes anomaly scores by summing attribute frequencies. Data points having features with low occurrence probability (estimated from frequencies) are likely to be outliers. 
Other existing methods mixing categorical and numerical, e.g. SmartSifter~\cite{smartsifter} and ODMAD~\cite{odmad}, could be applied to pure categorical data. ODMAD performs an initial off-line pattern mining stage, while SmartSifter is, to the best of our knowledge, the only previous unsupervised online algorithm for categorical data. However, it is unclear whether it can scale to a large number of attributes.   
\section{Itemset and Association Rule Mining}
\label{sec:arm}

In association rule mining (ARM)~\cite{Agrawal93}, data comes as a \emph{transaction database} ${\mathcal D}$ (as in Table~\ref{tab:formalcontext}) involving a universe of \emph{items}, or attributes, ${\mathcal A} =\{a_1,a_2,\dots, a_n\}$ (here named $a$ to $e$). A set of items $X \subseteq {\mathcal A}$ is called an \emph{itemset}. Below, for compactness reasons, itemsets are given in separator-less form. Then, a transaction, aka object, is a pair ($tid$, itemset) where $tid$ is a \emph{transaction identifier} taken from the set ${\mathcal O}= \{o_{1}, o_{2}, ..., o_{m}\}$ (aka set of objects). 
Next, a set of tids is called a \emph{tidset}. For historical reasons, we call ${\mathcal D}$ a \textit{context} and introduce it as $\times$-table (see Table~\ref{tab:formalcontext}).

The image of a tidset $Y$ from $\mathcal{O}$, $\iota(Y) = \bigcap \{Z | (j,Z) \in \mathcal{D}, j \in Y\}$ is made of the items shared by all the corresponding objects. Conversely, the image of an itemset $X$, \textit{aka} its \emph{support set}, comprises the tids of all objects for which $X$ is a subset of the respective itemset, $\tau(X)= \{j | (j,Z) \in \mathcal{D}, X \subseteq Z\}$. For instance, in Table~\ref{tab:formalcontext}, $\tau(e)=\{o_{1}\}$. An itemset quality reflects its support: (absolute) support is its supporting set size, $supp_a(X)=|\tau(X)|$ while relative support is the fraction $supp(X)=|\tau(X)|/|\mathcal{D}|$.
A support threshold, \textit{min\_supp}, splits itemsets into \textit{frequent} and infrequent itemsets, whereas with a \textit{max\_supp}, infrequent ones, a.k.a. \textit{rare}~\cite{Szathmary07}, are the target. 
For instance, with \textit{min\_supp}=3, $ac$ is frequent  while $abcd$ is rare.
Here, we exploit borderline cases,
i.e. \textit{maximal frequent itemsets} (\textit{MFI}, no frequent superset) and \textit{minimal rare itemsets} (\textit{MRI}, no rare subsets), which are closed itemsets and generators, respectively.
In fact, $\tau$ induces an equivalence relations on $\wp(\mathcal{A})$: $X \equiv Z$ iff $\tau(X) = \tau(Z)$ whereby each class has a unique maximum, called the \textit{closed} itemset, and one or more minima, its \textit{generators}. 
Both categories admit straightforward support-wise definitions:  
an itemset $X$ is closed (generator) if it admits no superset (subset) with identical support.
For instance, in Table~\ref{tab:formalcontext}, $bc$ is closed while $b$ is a (non closed) generator. 
\begin{table}
\centering
\scriptsize
\begin{tabular}{|l|c c c c c|}
\hline
 & \textbf{$a$} & \textbf{$b$} & \textbf{$c$} & \textbf{$d$} & \textbf{$e$} \\ \hline 
\textbf{$o_{1}$}                           & ~          & x          & x          & ~          & x          \\ 
\textbf{$o_{2}$}                           & x          & ~          & x          & x          & ~          \\ 
\textbf{$o_{3}$}                           & x          & x          & x          & x          & ~          \\ 
\textbf{$o_{4}$}                           & x          & ~          & ~          & x          & ~          \\ 
\textbf{$o_{5}$}                           & x          & x          & x          & x          & ~          \\ 
\textbf{$o_{6}$}                           & x          & ~          & x          & x          & ~          \\ \hline
\end{tabular}
\caption{A sample transaction database (context).}
\label{tab:formalcontext}
\vspace{-4mm}
\end{table}

An association rule has the form $X$$\rightarrow$$Y$, where $X, Y \subseteq \mathcal{A}$ and $X$ $\cap$ $Y$= $\emptyset$. 
Among the most popular rule quality measures are the \textit{support}, \textit{supp}($X$$\rightarrow$$Y$)=\textit{supp}($X$$\cup$$Y$), the \textit{confidence}, \textit{conf}($X$$\rightarrow$$Y$)=\textit{supp}($X$$\rightarrow$$Y$)/\textit{supp}(X),
and the \textit{lift}, \textit{lift}($X$$\rightarrow$$Y$)= \textit{supp}($X$$\cup$$Y$)/\textit{supp}($X$)$\times$\textit{supp}($Y$). 
Again, support-wise, a rule qualifies as \textit{frequent} or \textit{rare}~\cite{Szathmary07} whereas it is said to be \textit{confident} if it reaches a confidence threshold $min\_conf$.  Frequent (rare) and confident rules are \textit{valid}.
For example, with \textit{min\_conf}= 3/5, 
and \textit{min\_supp}=\textit{max\_supp}=3,
$a$$\rightarrow$$c$ is valid frequent (\textit{supp}= 4 and \textit{conf}=4/5) while 
$ab$$\rightarrow$$d$ is valid rare (\textit{supp}=2 and \textit{conf}=1).

\section{Rule Mining-Based Anomaly Detection}
\label{sec:approach}
Frequent and rare patterns~\cite{Szathmary07} clearly convey different types of regularities in the data. The former tend to capture generally valid rules, e.g. the typical behavior of customers on an online retail portal.
The latter, in contrast, focus on local regularities that, without being totally exceptional, i.e. unique, have a very limited extent. In fact, such regularities will likely be missed by an exhaustive search with a low $min\_supp$ threshold since they will be missed in the immensely voluminous result. Rare patterns have the capacity to capture the features of a highly under-represented --- and unknown --- class in an imbalanced sample, which makes them valuable for our study. 

Following the above guidelines, we hypothesize that APT-related processes constitute a separate class following a event chaining schema that distinguishes them from normal ones. The schema should manifest as atypical associations of events which, taken separately, might be totally legitimate. Consequently, such atypical behavior may be captured either as a mismatch to common behavioral patterns or as a consistent, yet rare, pattern of itself. Thus, a forensic APT detection approach should focus on candidate processes satisfying these conditions.
Technically speaking, an anomaly score needs to quantify the above (mis)matches of key patterns while also reflecting those patterns' quality (see section~\ref{sec:arm}). 

As a first approach, we propose to organize system traces as a transaction database and to mine itemset associations from them (rather than, say, sequential patterns). Such an approach has the added benefit of being explainable as the event implication format is easy to interpret. This, in turn, favors better understanding of mechanisms behind an APT and therefore should facilitate threat prevention.

\subsection{Anomaly Detection Method}
\label{sec:method}
We designed two separate anomaly detection methods: \textit{VR-ARM} for Valid Rare ARM and \textit{VF-ARM} for Valid Frequent ARM. The underlying association rules have been derived from \textit{MRIs} and \textit{MFIs}, respectively. The advantage behind using both
here is to 
speed up the ARM task w.r.t. other direct approaches~\cite{SzathmaryVNG12} due to a substantial reduction in the underlying search space. 
Algorithm~\ref{alg:anomalydet} presents the pseudo-code of VR-ARM. It takes as input a context $C$ (i.e. a $m \times n$ $\times$-table) plus the support and confidence thresholds. Its outputs are (i) $Rules$: the association rules extracted from $C$; (ii) $Attacks$: the list of anomalous objects; and (iii) $Scores$: the list of scores associated to objects in $Attacks$. The algorithm starts by calling $GetRareRules$ to generate the rare rules through the \textit{MRIs}. It implements the \textit{BtB} (\textit{Break the Barrier}) method~\cite{SzathmaryVNG12}. Next, each object from $C$ is matched against the rare associations: an object is said to satisfy a rare rule if the rule's itemsets are included in the object's  itemsets. A score is assigned based on the set of matching outcomes, which represents the degree of anomalousness of the matched object (the higher the score, the more anomalous the object).
VF-ARM follows the same principles as VR-ARM so will be skipped here: main differences include the parameter $min\_supp$ and a call to the MFI-based rule miner $GetFreqRules$. Moreover, an object violates a frequent rule if its itemsets are included in the left-hand side of the rule and not in its right-hand side.
Objects matching rare rules or violating frequent ones are deemed potentially anomalous, hence put in $Attacks$. Their anomalousness scores are computed as the average of a function combining the interestingness (e.g. lift) and length of the respective rules, $log_{2}(1-Interest(R[j]))*Length(R[j])$. High quality rules, with many items on both sides, would have large scores.

At a final stage, the list of the potential attack objects are ranked w.r.t. their anomaly scores so that the top-ranked elements could be subsequently checked by a security expert. 
\begin{algorithm}
\SetAlgoLined
\scriptsize
  \caption{VR-ARM: Rare Rule mining anomaly detection.}
  \label{alg:anomalydet}
   \SetKwInOut{Input}{inputs}
  \SetKwInOut{Output}{outputs}
\SetKwBlock{Beginn}{beginn}{ende}
    \Input{A context $C[m,n]$, $max\_supp$, $min\_conf$}
    \Output{ $Rules[0:r]$; // \textit{A list of association rules} \\
             $Attacks[0:a]_{a \leq m}$; //\textit{A list of anomalous objects} \\
             $Scores[0:a]_{a \leq m}$; // \textit{A list of attack scores}
             }
\Begin{           
$Rules$ $=$
GetRareRules($C$,$max\_supp$, $min\_conf$)\;
    \ForEach{Object $O[i]_{1 \leq i \leq m}$ in $C$ }{%
    $Score[i]=0.0$\;
    $IsAttack=False$\;
     \ForEach{Rule $R[j]$ in $Rules$ }{
         
       \If{$O[i]$ satisfies $R[j]$}{
       $IsAttack=True$\;
        $Score[i]=Score[i]+|log_{2}(1-Interest(R[j]))*Length(R[j])|$
               }
    
    %
    }
             \If{IsAttack==True}{%
                     $Append(Attacks,O_{i})$\;
                     $Append(Scores,Score[i])$\;
                                        }
    }
    Rank($Attacks$ using $Scores$)\;
    \KwRet{$(Rules,Attacks,Scores)$}\;
}
 \end{algorithm}
\setlength{\textfloatsep}{03pt}


\section{Results and Discussion}
\label{sec:results}

\subsection{Datasets}
\label{subsec:datasets}
For our evaluation, we use two data collections described in ~\cite{DBLP:journals/fgcs/BerradaCBMMTW20}, publicly available\footnote{\url{https://gitlab.com/adaptdata}}. These collections (or scenarios) consist of sets of feature views or \textit{contexts} from raw whole-system provenance graphs produced during two DARPA Transparent Computing (TC) program ``engagements'' (exercises aimed at evaluating provenance recorders and techniques to detect APT activity from provenance data). During the engagements, several days (5 for scenario 1 and 8 for scenario 2) of system activity (process, netflow, etc.) were recorded on various platforms (Windows, BSD, Linux and Android) subjected to APT-like attacks. \cite{unicorn} explains engagements in more detail and provides more information on provenance data for Scenario 2 (referred to as Engagement 3).
All \textit{contexts} in the data collections relate unique process identifiers (rows) to varying features/attributes of process activity: types of events performed by process ($ProcessEvent$ (PE)), name of process executable ($ProcessExec$ (PX)), name of parent process executable ($ProcessParent$ (PP)), IP addresses and ports accessed by process ($ProcessNetflow$ (PN)), joining all the previous features (renamed, if needed, to avoid ambiguity) together ($ProcessAll$ (PA)). Table~\ref{datatable} summarizes the properties of all contexts per collection/scenario. For a more detailed description,
see~\cite{DBLP:journals/fgcs/BerradaCBMMTW20}. We might observe that \textbf{(a):} The number of processes varies widely: Linux has 3 to 10-times more that Windows/BSD and up to 2400 times more than Android.  \textbf{(b):} The number of processes/attributes is unrelated to the original dataset size. Thus, albeit the largest, the Android dataset has the fewest processes and attributes (due to the provenance recorder
mostly logging low-level app activity and performing dynamic information flow tracking, which are pieces of information we do not analyze).

\begin{table*}
\centering
	\tiny
\resizebox{0.99\textwidth}{!}{
\begin{tabular}{|l|l||l|l|l|l|l|l|l|l|}
\hline & Scenario & Size& $PE$   & $PX$  & $PP$  & $PN$     & $PA$  & $nb\_attacks$    & $\%\frac{nb\_attacks}{nb\_processes}$     \\ \hline \hline
BSD    & 1 &288 MB &76903 / 29  & 76698 / 107  & 76455 / 24  & 31 / 136  & 76903 / 296 & 13&0.02\\  
    & 2 &1.27 GB &224624 / 31  &224246 / 135  & 223780 / 37  & 42888 / 62 &  224624 / 265      & 11&0.004\\ \hline
Windows & 1 &743 MB & 17569 / 22    &  17552 / 215  &   14007 / 77        &   92 / 13963      & 17569 / 14431& 8&0.04\\  
   & 2 &9.53 GB& 11151 / 30    &  11077 / 388  & 10922 / 84  & 329 / 125      &  11151 / 606    &8&0.07\\ \hline
Linux  & 1 &2858 MB &247160 / 24 & 186726 / 154 & 173211 / 40 & 3125 / 81 & 247160 / 299  &25&0.01\\
    & 2 &25.9 GB &282087 / 25 & 271088 / 140 & 263730 / 45 &6589 / 6225 &  282104 / 6435      &46&0.01\\ \hline
Android& 1 &2688 MB&102 / 21     &102 / 42&0 / 0&8 / 17& 102 / 80&9&8.8\\
&2 &10.9 GB&12106 / 27     &12106 / 44&24 / 11&4550 / 213&12106 / 295 &13&0.10\\ \hline
\end{tabular}
}

\caption{Experimental dataset metrics. A context entry (columns 4 to 8) is number of rows (processes) / number of columns (attributes).}
 \label{datatable}
\end{table*}

\subsection{Evaluation Metrics}
Rather than classifying processes as anomalous or not, our method  ranks them w.r.t. their degree of suspiciousness.
Due to the high data imbalance (\textit{attacks} amount to 0.004\% to 8.8\% of data), using classification with accuracy would lead to the \textit{accuracy paradox}~\cite{thomas2008}. 

Instead, a better suited metric would be \textit{normalized discounted gain} (nDCG), first introduced in~\cite{jarvelin2002cumulated} as a means to assess ranking quality in information retrieval. It is intended to reflect the value the end user assigns to a ranked document, i.e. relevant documents are more valuable than marginally relevant ones and even more so than irrelevant ones. Also, intuitively, since a user will only scan a small portion of a ranking, relevant documents close to its top have more value than comparably relevant ones further down the list (those too far down would be skipped). Consequently, nDCG favors rankings with all relevant documents near the top. 
The same principle applies to anomalous process: low ranked attack processes are all but useless to analysts whose monitoring burden increases with the amount of events to inspect (up to a point where suspicious processes escape their attention). Also, a large number of top-ranked normal processes (false alarms) may degrade analyst confidence in the automated monitoring system.
The \textit{discounted cumulative gain} (DCG) of a ranking sums element relevance scores penalized by the logarithm of their rank:

$DCG_N=\sum_{i=1}^N\frac{rel_i}{\log_2(i+1)}$, where $N$ is the ranking size, $rel_i$ the $i$-th element relevance score. As DCG scores are not comparable for lists of varying size, a normalization is performed with the ideal score iDCG, i.e. one corresponding to the best case (all relevant entities at the top). Thus, for a ranking with $p$
relevant entities $iDCG_N =\sum_{i=1}^p\frac{rel_i}{\log_2(i+1)}$. nDCG is the normalized version of DCG: $ nDCG_N =\frac{DCG_N}{iDCG_N}$.
We use binary scale for relevance $rel_i$:
1 for attack processes, 0 for normal ones.

\subsection{Evaluation Results}
\subsubsection{APT Ranking Visualisation with Band Diagrams}
\begin{figure}[h!]
\centering
\includegraphics[width=85mm,height=50.35mm]{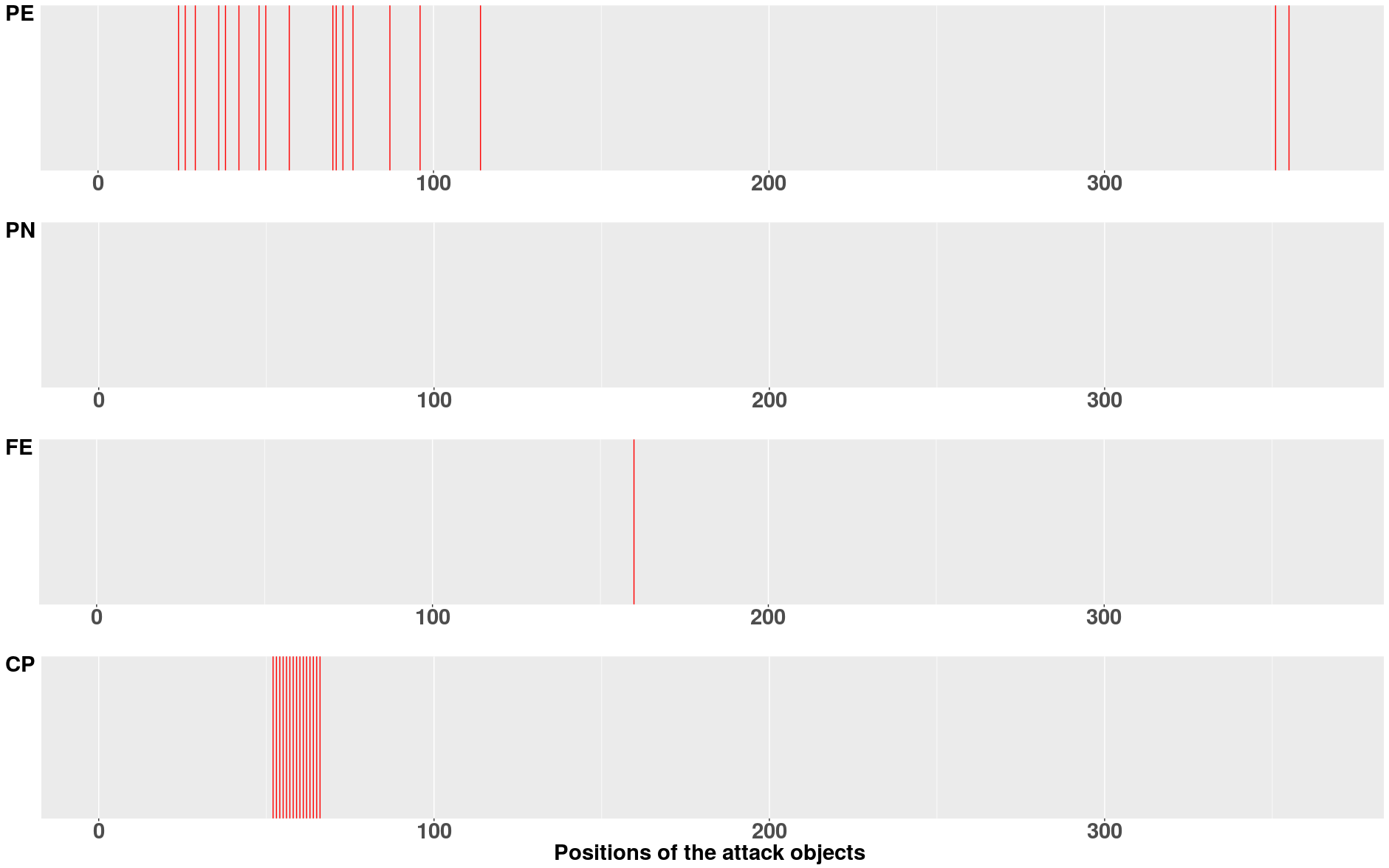}
\caption{Band diagrams representing the positions of the attacks (true positives) in some contexts of the BSD dataset (scenario 1). The x-axis represents the attack positions in ranked lists.}
\label{fig:band}
\end{figure}

To ease the inspection of generated rankings, we designed a visualization technique, called \textit{band diagram charts} (see figure~\ref{fig:band}). Each ranked list is drawn as a horizontal band, where the position of a true positive (APT) is marked by red lines. Top ranked entities will be put on the left, thus multiple red lines in that area reveal highly efficient anomaly detection. 

\subsubsection{Ranking Evaluation With nDCG}
We compared VF-ARM and VR-ARM to existing tools such as FPOutlier (FPOF)~\cite{fpoutlier}, Outlier-degree (OD)~\cite{outlierdegree},  CompreX~\cite{comprexpaper}, AVF (Attribute Value Frequency)~\cite{avf} and OC3 (Krimp)~\cite{krimp}: We ran them on the contexts from Table~\ref{datatable} and assessed the resulting rankings by means of nDCG.
To that end, we first reimplemented FPOF, AVF, OC3 and OD in Python following their original descriptions\footnote{Code available at \url{https://gitlab.com/anomalyDetection/baseline}}.  We reused publicly-available implementations of OC3~\cite{krimp-ad} and CompreX \cite{comprexpaper}, in C++ and Matlab respectively. 
Finally, experiments were run on an Intel Core i7-6700 CPU
(3.4 GHz), 32 GB RAM PC with Ubuntu OS. 

\begin{table*}[!htb]

\centering
	\tiny
	\begin{adjustbox}{max width=1\textwidth,center}
\begin{tabular}{|l|l||c|c|c|c|c|c|c|c|c|c||c|c|c|c|}
\hline
\multirow{2}{*}{}                & \textbf{Detection method} & \multicolumn{2}{c|}{FPOF} & \multicolumn{2}{c|}{OD} & \multicolumn{2}{c|}{Comprex}& \multicolumn{2}{c|}{OC3} & \multicolumn{2}{c||}{AVF}  & \multicolumn{2}{c|}{VR-ARM (sup x conf)} & \multicolumn{2}{c|}{VF-ARM (sup x conf)}  \\ \cline{2-16} 
                                 & \textbf{Attack scenario}  & 1           & 2     & 1            & 2 & 1            & 2       & 1          & 2          & 1             & 2            & 1            & 2            & 1            & 2            \\ \hline \hline
\multirow{4}{*}{\textbf{\rotatebox{90}{Source}}} & \textit{Windows}          &  0.20            &    DNF         &    0.20          &    DNF        &       0.60           &       DNF    &0.30&\textbf{0.23}&0.60&0.21   &       \colorbox{red}{\textbf{0.82} (5x100)}       &     0.19 (5x100)         &  0.33  (60x70)             &        0.13 (30x30)      \\ \cline{2-16} 
                                 & \textit{BSD}            & 0.20             &     0.13        &    0.19          &0.17            &    0.54              &       DNF     &0.43&\textbf{0.24}&0.51&0.19    &      \colorbox{orange}{\textbf{0.64} (0.05x100)}          &   0.12 (5x100)           &   0.33 (40x97)            &      0.12 (40x40)        \\ \cline{2-16} 
                                 & \textit{Linux}              &   0.18            &     0.22        &    0.18        &   0.21         &   0.30               &      DNF   &\textbf{0.38}&\textbf{0.38}&0.27&0.29       &    0.13 (0.05x100)            &      0.14 (5x100)        &        0.22 (20x97)        &        0.10 (40x40)      \\ \cline{2-16} 
                                 & \textit{Android}          &     0.29        &    0.36         &   0.33          &      0.22     &    0.82             &   DNF        &0.74&0.32&0.84&0.30       &     \colorbox{red}{\textbf{0.87} (30x100)}          &     \colorbox{orange}{\textbf{0.50} (5x100)}          &    0.77 (5x70)             &   0.12 (5x70)            \\ \hline
\end{tabular}
\end{adjustbox}
\caption{Evaluation of anomaly scoring: ProcessEvent (PE)}
\label{tab:batch-evaluation-pe} 


\centering
	\begin{adjustbox}{max width=1\textwidth,center}
\begin{tabular}{|l|l||c|c|c|c|c|c|c|c|c|c||c|c|c|c|}
\hline
\multirow{2}{*}{}                & \textbf{Detection method} & \multicolumn{2}{c|}{FPOF} & \multicolumn{2}{c|}{OD} & \multicolumn{2}{c|}{Comprex}& \multicolumn{2}{c|}{OC3} & \multicolumn{2}{c||}{AVF}  & \multicolumn{2}{c|}{VR-ARM (sup x conf)} & \multicolumn{2}{c|}{VF-ARM (sup x conf)}  \\ \cline{2-16} 
                                 & \textbf{Attack scenario}  & 1           & 2     & 1            & 2 & 1            & 2       & 1          & 2          & 1             & 2            & 1            & 2            & 1            & 2            \\ \hline \hline
\multirow{4}{*}{\textbf{\rotatebox{90}{Source}}} & \textit{Windows}          &  0.15               &   DNF          &    0.15        &    DNF        &   DNF        &       DNF        &\textbf{ 0.28}&0.24&\textbf{ 0.28}&\textbf{0.22}  &   0    &      0    &         0      &         0     \\ \cline{2-16} 
                                 & \textit{BSD}            & 0.15              &       0.18      &    0.15         &0.17            &    DNF             &     DNF        &0.49&\textbf{0.51}&0.34&0.17   &      0.08 (0.05x100)             & 0.05 (0.05x100)          &   \colorbox{orange}{\textbf{0.53} (0.001x10)}             & 0.06 (1x40)      \\ \cline{2-16} 
                                 & \textit{Linux}              &     0.18          & 0.20             &   0.18       & 0.20           &   DNF              &       DNF     &0.30    &0.42&\textbf{ 0.43}&\textbf{0.42}& 0.12 (0.05x100)  &  0         &           0.10 (0.001x0.1)       &       0.004 (0.001x0.1)\\ \cline{2-16} 
                                 & \textit{Android}          &    0.22             &      0.29       &  0.22          &     0.29       &    0.22             &     DNF    &\textbf{ 0.39}&\textbf{ 0.39}&  0.39 &0.38       &    0       &    0     & 0    &  0          \\ \hline
\end{tabular}
\end{adjustbox}
\caption{Evaluation of anomaly scoring: ProcessExec (PX)}
\label{tab:batch-evaluation-px} 


\centering
	\begin{adjustbox}{max width=1\textwidth,center}
\begin{tabular}{|l|l||c|c|c|c|c|c|c|c|c|c||c|c|c|c|}
\hline
\multirow{2}{*}{}                & \textbf{Detection method} & \multicolumn{2}{c|}{FPOF} & \multicolumn{2}{c|}{OD} & \multicolumn{2}{c|}{Comprex}& \multicolumn{2}{c|}{OC3} & \multicolumn{2}{c||}{AVF}  & \multicolumn{2}{c|}{VR-ARM (sup x conf)} & \multicolumn{2}{c|}{VF-ARM (sup x conf)}  \\ \cline{2-16} 
                                 & \textbf{Attack scenario}  & 1           & 2     & 1            & 2 & 1            & 2       & 1          & 2          & 1             & 2            & 1            & 2            & 1            & 2            \\ \hline \hline
\multirow{3}{*}{\textbf{\rotatebox{90}{Source}}} & \textit{Windows}          &    0.10            & DNF            &   0.10         &    DNF        &      DNF     &        DNF    &\textbf{0.21}&\textbf{0.22}&\textbf{0.21}&\textbf{0.22}   &   0   &  0         &     0            &       0       \\ \cline{2-16} 
                                 & \textit{BSD}            &    0.13         &         0.10    &           0.13     &0.09            & DNF             &    DNF     &0.43&\textbf{0.29}&0.30&0.17      &      0.29 (0.05x100)&     0.24 (0.05x100)   &         \colorbox{orange}{\textbf{0.67} (0.001x10)}     &       0.06 (0.001x10)    \\ \cline{2-16} 
                                 & \textit{Linux}              &     0.17         &     0.20        &   0.17       &    0.20        &   DNF         &    DNF      &\textbf{0.24}&\textbf{0.42}&0.20  &0.25   &    0          &        0   &     0.12 (0.001x1)           &        0.03 (10x40)      \\ \cline{2-16} 
                            \hline
\end{tabular}
 \end{adjustbox}
\caption{Evaluation of anomaly scoring: ProcessParent (PP)}
\label{tab:batch-evaluation-pp}

\centering
\begin{adjustbox}{max width=1\textwidth,center}
\begin{tabular}{|l|l||c|c|c|c|c|c|c|c|c|c||c|c|c|c|}
\hline
\multirow{2}{*}{}                & \textbf{Detection method} & \multicolumn{2}{c|}{FPOF} & \multicolumn{2}{c|}{OD} & \multicolumn{2}{c|}{Comprex}& \multicolumn{2}{c|}{OC3} & \multicolumn{2}{c||}{AVF}  & \multicolumn{2}{c|}{VR-ARM (sup x conf)} & \multicolumn{2}{c|}{VF-ARM (sup x conf)}  \\ \cline{2-16} 
                                 & \textbf{Attack scenario}  & 1           & 2     & 1            & 2 & 1            & 2       & 1          & 2          & 1             & 2            & 1            & 2            & 1            & 2            \\ \hline \hline
\multirow{4}{*}{\textbf{\rotatebox{90}{Source}}} & \textit{Windows}          &         0.36       &DNF             &     0.36       &  DNF          &  DNF         &     DNF     &\textbf{0.65}&0.24&0.58&0.18      &  \colorbox{orange}{\textbf{0.62} (10x100)}       &    \colorbox{green}{\textbf{0.30} (5x100)}  &    0           &       0       \\ \cline{2-16} 
                                 & \textit{BSD}            &        0.13       &       0.20      &      0.14        &       0.20     & DNF                &          DNF   &0.11&0.50&\textbf{0.58}&0.18   &     0.34  (10x100)              &   \colorbox{orange}{\textbf{0.60} (5x100)}      &   0.11 (5x60)              &     0.18 (5x60)      \\ \cline{2-16} 
                                 & \textit{Linux}              &  0.23             &   0.31          &   0.23          &    0.23        &   DNF              &        DNF     &0.38&0.35&0.31&\textbf{0.42}   &     \colorbox{orange}{\textbf{0.58} (20x100)}           & 0.39 (5x100)          &   0.42 (5x50)                   &         0.11 (10x40)     \\ \cline{2-16} 
                                 & \textit{Android}          &   0.42          &    0.36         &   0.36         & 0.34           &   DNF              &      DNF      &\textbf{0.64}&0.30&0.47&0.32    &    0.46  (50x100)              &  \colorbox{orange}{\textbf{0.71} (30x100)}         &   0  &  0.10 (20x50)          \\ \hline
\end{tabular}
 \end{adjustbox}
\caption{Evaluation of anomaly scoring: ProcessNetflow (PN)}
\label{tab:batch-evaluation-pn} 


\centering
	\begin{adjustbox}{max width=1\textwidth,center} 
\begin{tabular}{|l|l||c|c|c|c|c|c|c|c|c|c||c|c|c|c|}
\hline
\multirow{2}{*}{}                & \textbf{Detection method} & \multicolumn{2}{c|}{FPOF} & \multicolumn{2}{c|}{OD} & \multicolumn{2}{c|}{Comprex}& \multicolumn{2}{c|}{OC3} & \multicolumn{2}{c||}{AVF}  & \multicolumn{2}{c|}{VR-ARM (sup x conf)} & \multicolumn{2}{c|}{VF-ARM (sup x conf)}  \\ \cline{2-16} 
                                 & \textbf{Attack scenario}  & 1           & 2     & 1            & 2 & 1            & 2       & 1          & 2          & 1             & 2            & 1            & 2            & 1            & 2            \\ \hline \hline
\multirow{4}{*}{\textbf{\rotatebox{90}{Source}}} & \textit{Windows}          &      DNF        &     DNF        &  DNF        &DNF&     DNF      &       DNF      &0.49&DNF&0.52&DNF   &  \colorbox{orange}{\textbf{0.61} (5x100)}      &    \colorbox{green}{\textbf{0.35} (5x100) }      &   0.50 (80x70)              &      0.07 (40x40)        \\ \cline{2-16} 
                                 & \textit{BSD}            &      0.21      &   0.15          &0.19 &0.15            &  DNF               &    DNF      &0.38&DNF&\textbf{0.52}&DNF      &    0.36 (0.05x100)               &    \colorbox{orange}{\textbf{0.52} (5x100) }      &  0.18 (97x97)              &      0.14 (40x40)     \\ \cline{2-16} 
                                 & \textit{Linux}              &   0.18           &    DNF         &     0.18       &  DNF          &DNF                 &   DNF       &0.41&DNF&0.29&DNF      &   \colorbox{orange}{\textbf{0.54} (0.05x100)}                &     \colorbox{green}{\textbf{0.45} (5x100) }     &       0.13 (40x70)           &      0.09 (40x40)        \\ \cline{2-16} 
                                 & \textit{Android}          &   0.31          &         0.37    &   0.34            &          0.20  &  DNF               &          DNF    & 0.82&0.40& \textbf{0.83}&0.35  &    0           & \colorbox{orange}{\textbf{0.51} (0.05x100)}       &0    &           0.43 (40x40) \\ \hline
\end{tabular}
\end{adjustbox}
\caption{Evaluation of anomaly scoring: ProcessAll (PA)}
\label{tab:batch-evaluation-pa} 

 \end{table*} 

The global set of nDCG scores is split context-wise into tables \ref{tab:batch-evaluation-pe}, \ref{tab:batch-evaluation-px}, \ref{tab:batch-evaluation-pp}, \ref{tab:batch-evaluation-pn}, and \ref{tab:batch-evaluation-pa}. An entry here corresponds to a combination (method, scenario, OS). Noteworthily, some algorithms did not finish within a reasonable time (4 to 48 hours). Such cases are indicated by $DNF$.
For each OS (row) vs scenario combination, the maximum nDCG score is marked in bold. For visibility, high scores of our approach, VR-ARM or VF-ARM, are colored: green for values in ]0.25,50], orange for ]0.50, 0.75] and red for top scores in ]0.75, 1].
 
VR-ARM and OC3 were competitive on the $ProcessEvent$ context with considerably better scores obtained with VR-ARM between 0.50 (Android, scenario 2) and up to 0.87 (Android, scenario 1). FPOF and OD yielded similar scores to each other. AVF and VF-ARM  produced good scores with scenario 1. Concerning $ProcessExec$ and $ProcessParent$, AVF and OC3 generated moderately acceptable scores with all four OS. The highest scores in these contexts were obtained by OC3 (0.51 and 0.42) in BSD w.r.t. scenario 2. VF-ARM got good ranking scores for BSD (scenario 1) for both contexts, yet due to the low confidence values, they were deemed irrelevant and non competitive with the other methods on this task. Concerning attacks in $ProcessNetflow$, VR-ARM reached good results on both attack scenarios, with nDCG scores varying between 0.30 (Windows, scenario 2) and 0.71 (Android, scenario 2).
Like-wise, top-ranked anomalous entities in the super-context $ProcessAll$ have been detected with VR-ARM on both attack scenarios. AVF and OC3 have also generated good scores with attack scenario 1. Note that CompreX produced  good results whenever it was able to finish within a reasonable time ($ProcessEvent$, scenario 1); for wider contexts such as $ProcessExec$ or $ProcessParent$ or $ProcessAll$, it usually did not terminate within a few minutes (while~\cite{comprexpaper} claims CompreX can be run in anytime-mode, the available Matlab implementation does not support it).  Runtime-wise, FPOF, CompreX and OD were significantly more expensive than VR-ARM and AVF.

\begin{figure*}[!htb]
\centering
\includegraphics[width= \textwidth, height=5cm]{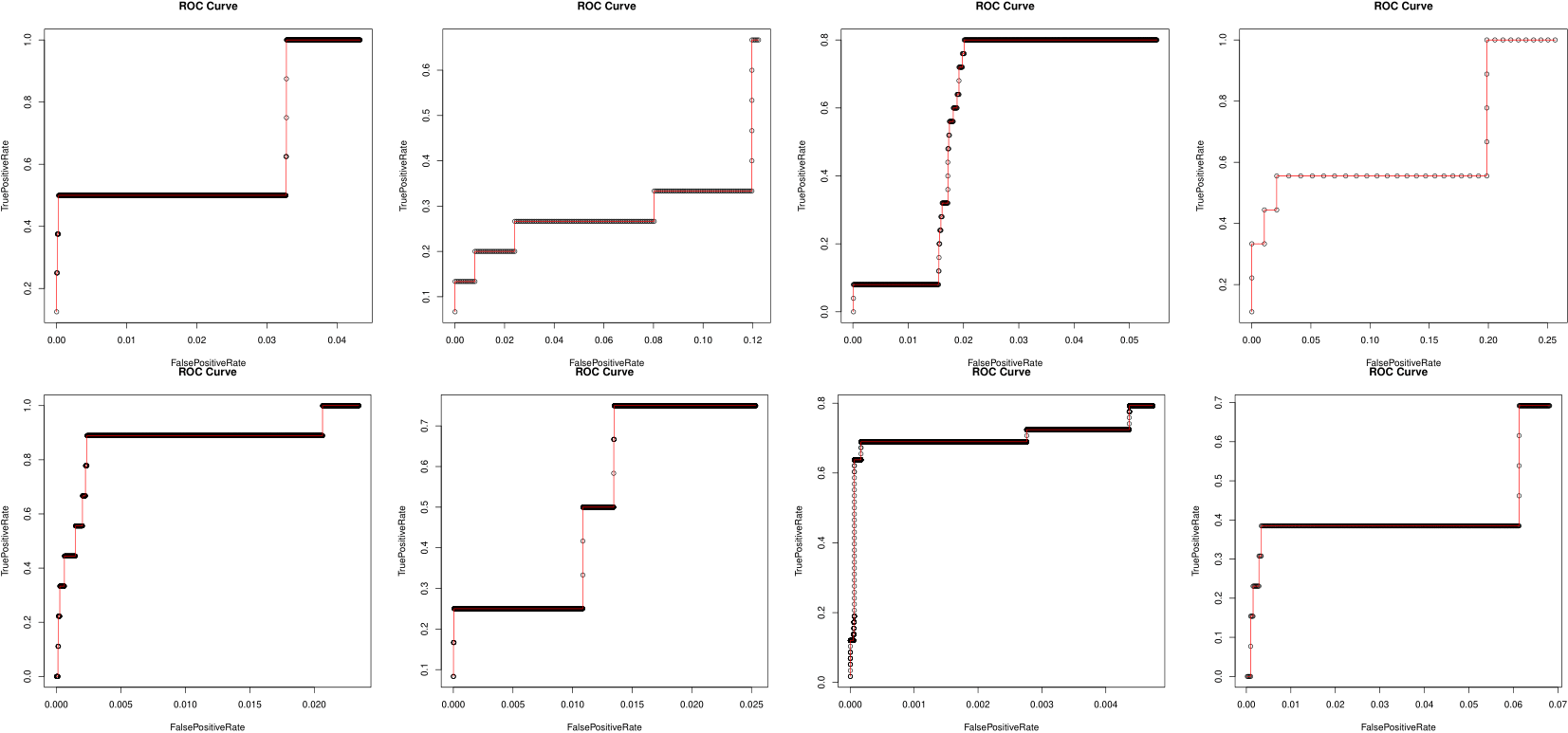}
\caption{ROC curves of the best anomaly scores obtained with the winner contexts (cf Table \ref{tab:winnercontext}). Top: scenario 1, bottom: scenario 2. Left to right: Windows, BSD, Linux, Android (the ROC curves use different scales in both axes due to space limits).}
\label{fig:roc}
\end{figure*}

Table~\ref{tab:winnercontext} summarizes attack detection and ranking for each combination (OS, scenario): The winner method for each OS is given with the best input context. Results comprise the top nDCG scores, the running time and the Area Under Curve (AUC). As per the figures, VR-ARM is likely to be the method that has led to the best nDCG scores with interesting AUC values on the four OS w.r.t. both scenarios (see also Figure~\ref{fig:roc}). Its time efficiency is arguably due to MRI-based associations being only a tiny fraction of all the rules.  

Intuitively, the results seem to confirm that most of the attacks can be detected by tracking their uncommon (rare) behavior in the provenance data. Indeed, in our experiments, we ran VR-ARM with \textit{max\_supp} values ranging between 0.05 and 30 \%. More interestingly, the rare rules found have confidence of 100\%. Note that we  kept the best configuration for every method that needs parameter tuning. Here again, VR-ARM stood out as very competitive even when compared to the best configurations of its competitors.

\begin{table*}[!htb]
 	\scriptsize
  	\scriptsize
  		\begin{adjustbox}{max width=1\textwidth,center} 
\begin{tabular}{|l|l|c|c|c|c|c|c|c|c|c|c|}
\hline
\multicolumn{2}{|l|}{}                                      & \multicolumn{2}{c|}{\textbf{Winner AD method}} & \multicolumn{2}{c|}{\textbf{Winner context}} & \multicolumn{2}{c|}{\textbf{Running time (sec.)}} & \multicolumn{2}{c|}{\textbf{nDCG}}  & \multicolumn{2}{c|}{\textbf{AUC}}\\ \hline  \hline

                                 & \textbf{Attack scenario} & 1                      & 2                     & 1                     & 2                    & 1               & 2               & 1                & 2   & 1                & 2               \\ \hline
\multirow{4}{*}{\textbf{Source}} & \textit{Windows}         & VR-ARM                 & VR-ARM                & ProcessEvent          & ProcessAll           &  4.60            &     23.86        & 0.82             & 0.35    &0.75            & 0.50         \\ \cline{2-12} 
                                 & \textit{BSD}             & VR-ARM                 & VR-ARM                & ProcessEvent          & ProcessNetflow       &    12.18           &  12.30          & 0.64             & 0.60          & 0.75            & 0.50  \\ \cline{2-12} 
                                 & \textit{Linux}           & VR-ARM                 & VR-ARM                & ProcessNetflow        & ProcessAll           &  3.53           &    2.74         & 0.58             & 0.45           & 0.83            & 0.50  \\ \cline{2-12} 
                                 & \textit{Android}         & VR-ARM                 & VR-ARM                & ProcessEvent          & ProcessNetflow       &    0.78         &   3.35         & 0.87             & 0.71          &0.92            & 0.50   \\ \hline
\end{tabular}
 \end{adjustbox}
  \caption{Highest AUC and nDCG scores of the rule-mining anomaly detection (AD) methods for each database.
 }
   \label{tab:winnercontext}
\end{table*}

Note that our tool provides an explainable output, e.g. when tracing back a process with anomalousness score of 79.0 (highest), one of the rules it satisfied comprised rare itemsets of Events (OPEN, READ, CHANGEPRINCIPAL, SENDMSG), of Network activities (216.66.26.25:1025', '216.66.26.25:0', '128.55.12.10:53'), and of File activities ('/etc/host.conf','/lib/x86\_64\_linux\_gnu/libnss\_dns.so'). After interpretation, the rule expressed the fact that matching processes alter the library files and try to communicate with the listed ip addresses and subsequently send or receive data.
\section{Conclusion}
\label{sec:concl}
We proposed a novel OS-agnostic APT detection method which detects
anomalous entities using an association rule-based scoring and ranking,
and dedicated techniques visualizing potential anomalies within the rankings. It was evaluated on large datasets containing realistic APT on a variety of OS.
The rare rule version VR-ARM was shown to consistently rank attack processes as highly anomalous entities. A key advantage thereof is the easy interpretation of its results.

We are studying a stream version of the method to maintain MFI/MRI over a stream window (e.g. as in~\cite{martin2020ciclad}). We'll next examine the integration of analysts' feedback in the scoring loop as an active learning task. 
\subsection*{Acknowledgements}
Reported work was partially supported by 
DARPA
under contract FA8650-15-C-7557, by ERC grant Skye (grant  682315), and by an ISCF Metrology Fellowship grant provided by the UK Department for Business, Energy and Industrial Strategy. 
We thank H. Mookherjee for reimplementing FPOF and OD.
\bibliographystyle{named}

\bibliography{references2}
\end{document}